\documentclass[preprint,aps]{revtex4}
\usepackage{graphicx}

\def\dmu  {\mbox{$\downarrow$\hspace{-.2mm}$\mu$}} 
\def\dnu  {\mbox{$\downarrow$\hspace{-.2mm}$\nu_{\mu}$}} 
\def\umu  {\mbox{$\uparrow$\hspace{-.2mm}$\mu$}} 
\def\unu  {\mbox{$\uparrow$\hspace{-.2mm}$\nu_{\mu}$}}  

\begin{document}

\title{Vetoing atmospheric neutrinos in a high energy neutrino telescope}  
\author{Stefan Sch\"onert,$^1$ Thomas K. Gaisser,$^2$ Elisa Resconi,$^1$ Olaf Schulz$^1$} 
\affiliation{$^1$Max-Planck-Institut f\"ur Kernphysik \\ 
Saupfercheckweg 1, 69117 Heidelberg, Germany \\
$^2$Bartol Research Institute and Department of Physics \& Astronomy\\
University of Delaware, Newark, DE 19716 USA} 

\begin{abstract}   
We discuss the possibility to suppress downward atmospheric 
neutrinos in a high energy neutrino telescope. This can be achieved 
by vetoing the muon which is produced by the same parent meson decaying in the atmosphere.
In principle, atmospheric neutrinos with energies $E_\nu > 10$~TeV and zenith angle up to $60^\circ$
can be vetoed with an  efficiency of  $>99$\%.
Practical realization will depend on the depth of the neutrino telescope, on the 
muon veto efficiency and on the ability to
identify downward moving neutrinos with a good energy estimation.   
\end{abstract}  

\maketitle  
Neutrino telescopes such as IceCube at the South Pole \cite{icecube}, 
ANTARES in the Mediterranean Sea \cite{antares} and the Lake Baikal 
detector \cite{baikal} search
for neutrinos of extraterrestrial origin by using the Earth
as a filter to suppress the background of
atmospheric muons.  The charged current $\nu_\mu$ channel
is favored at TeV to PeV energies because of the large
muon range, up to several kilometers of matter.   
To prevent downward-going atmospheric muons (\dmu ) from being mis-identified as muons 
induced by neutrino interactions, neutrino telescopes typically search 
for neutrino-induced upward or horizontal muons (\unu  -\umu ) 
(Markov and Zheleznykh \cite{first}).  

The principle of detecting neutrinos by looking for \unu\ 
implies that the field of view of a neutrino telescope is limited to half of the sky 
i.e. the opposite hemisphere with respect to  the geographical position of the detector.  
Moreover, with this approach, atmospheric neutrinos that penetrate through the entire
Earth become an irreducible background for the search of extraterrestrial neutrinos.
There are two standard approaches to separating
extraterrestrial from atmospheric neutrinos in the \unu\ 
sample.  One is based on seeing an excess of events in a particular
direction/time interval (point source search \cite{AMANDAps}),
while the other is based on the assumption that astrophysical neutrinos
have a harder spectrum than the atmospheric \unu\ background  
(diffuse search~\cite{AMANDAdiffuse}).

The new generation of neutrino telescopes like IceCube \cite{icecube}, to be completed 
in 2011, and the R\&D project KM3NeT \cite{km3net}, will operate instrumented volumes 
of about one km$^3$. Among the new opportunities offered by such large detectors, 
one feature seems especially interesting to us: 
the use of part of the instrumented volume as an active veto for \dmu. 
This opens the field of
view of neutrino telescopes to the hemisphere above the detector \cite{deepcore}.

At the energies involved in a neutrino telescope (TeV-PeV), 
the opening angle 
between the \dnu\ and the \dmu\  
produced
in an atmospheric meson decay is very small. This implies that an
atmospheric \dnu\ has a certain probability to arrive in the detector 
accompanied by its
partner \dmu.

In this letter,  we discuss the
conditions under which a \dmu-veto will also veto atmospheric \dnu's. This  
opens the opportunity to suppress what it is considered so far
the irreducible background in neutrino telescopes.


At high energy, neutrinos come predominantly from decay of 
charged kaons and pions.  
Charged pions decay with a probability of 99.99\% 
through the reaction $\pi^\pm \rightarrow \mu^\pm + \nu_\mu (\bar{\nu}_{\mu})$ and 
charged kaons with a probability of 63.4 \% 
via $K^\pm \rightarrow \mu^\pm +\nu_\mu (\bar{\nu}_{\mu})$. 
There are small contributions from other channels, but we start by analyzing the
simplest and most important case of pion and kaon decay and discuss the
other cases at the end.   

\begin{figure*}[htbp] 
\centering       
\includegraphics[width=0.7\textwidth]{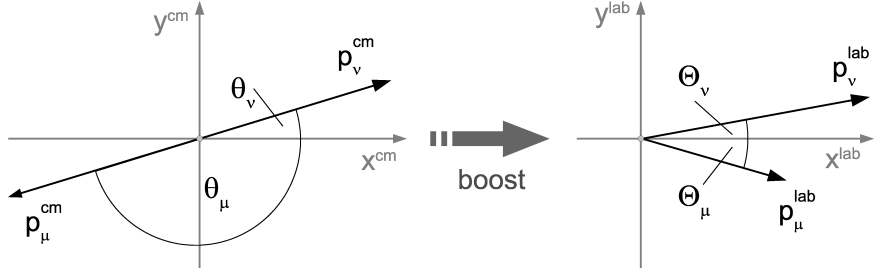}       
\caption{\small{Two body decay of the parent meson into muon and neutrino.       
The left figure displays the back-to-back kinematics in the meson       
center of mass (cm) frame. The right figure shows       
the momenta after Lorentz transformation into the laboratory frame.
 }}          
\label{fig:decay_kin} 
\end{figure*}  

Momentum conservation requires that the neutrino and companion muon are 
emitted back-to-back in the center of mass (cm) frame. The energies of the  muon 
and neutrino in this frame are 
\begin{eqnarray} E_\mu^{cm} &\simeq & \frac{m_i}{2} \,\, (1+r_i) \label{eq:e_cms_r} \nonumber \\
E_\nu^{cm} &\simeq &\frac{m_i}{2} \,\,(1-r_i)\,\,=|p^{cm}| 
\end{eqnarray}  
($c=1$).
Here $r_i = m_\mu^2/ m_i^2$ are the quadratic mass ratios, where 
the parent meson masses are $m_i$ ($i= \pi^\pm, K^\pm$), 
the muon mass is $m_\mu$ and we neglect the neutrino mass.   
Numerically, $r_\pi = 0.573$ and $r_K = 0.046$.     
Given the higher rest mass of the kaon with respect to that of the muon, 
the energy is quasi-equally shared between neutrino and muon for the 
kaon two-body decay, while for pion decay the energy balance is shifted more 
towards the muon. It should be noted that the muon energy is always larger 
than the neutrino energy in the cm-frame.  

Figure~\ref{fig:decay_kin} displays the kinematical relations in the  cm-frame 
and after Lorentz-transformation into the laboratory (lab) frame. 
After Lorentz-transformation along the positive x-axis from the cm- to 
the lab-frame, the muon and neutrino energies are given by:  
\begin{eqnarray} E_\nu = \gamma \, E_\nu^{cm} + \beta \, 
\gamma \, p_{x\nu}^{cm} \nonumber \\ 
E_\mu = \gamma \, E_\mu^{cm} + \beta \, \gamma \, p_{x\mu}^{cm}, 
\label{eq:emu_enu_lab} 
\end{eqnarray} 
where $\gamma$ and $\beta$ are the Lorentz factor and speed of the
parent pion or kaon. 
  
Taking into account the back-to-back emission of neutrino and muon  
in the cm-frame, 
\begin{eqnarray} 
\label{eq:momentacons}    
p_{x\nu}^{cm}& = &  |p^{cm}|\, \cos\theta_\nu \,\,\,\, {\rm and} \\ 
p_{x\mu}^{cm} & = &  
|p^{cm}|\, \cos(\theta_\nu - \pi) =  -|p^{cm}|\, \cos\theta_\nu , \nonumber  
\end{eqnarray} 
where $\theta_\nu$ is the angle of the neutrino in the cm-frame of
the parent meson relative to its direction in the lab-frame.
In the approximation $\beta\rightarrow 1$, which is valid
for meson energies above several GeV,
we can then rewrite Eq.~\ref{eq:emu_enu_lab} as
\begin{eqnarray}
E_\nu & = & \gamma|p^{cm}|(1+\cos\theta_\nu) \,\,\,\, {\rm and} \nonumber \\
E_\mu & = & \gamma|p^{cm}|\Big(\frac{1+r_i}{1-r_i} -\cos\theta_\nu\Big).
\label{eq:Eapprox}
\end{eqnarray}
The extreme condition for a given parent energy $E_i=E_\nu+E_\mu$ is
obtained for $\cos\theta_\nu=1$ for which $E_{\mu,min} = r_i E_{\nu,max}/(1-r_i)$.
Thus, for any $(E_i,E_\nu)$ the minimum 
energy of the companion muon is  
\begin{eqnarray} E_\mu & \ge & \Big(\frac{r_i}{1-r_i}\Big)\, E_\nu \,\,\,\, {\rm and} \nonumber \\
E_i & = & E_\mu\,+\,E_\nu\,\ge\,E_\nu\,\Big(\frac{1}{1-r_i}\Big)
\label{eq:mu_nu_lab} 
\end{eqnarray} with $i=\pi^\pm, K^\pm$. 
The corresponding numerical values for pion and kaon decays are 
\begin{eqnarray} \label{pions}
\pi^\pm: \quad \quad E_\pi \ge 2.342 \cdot E_\nu \\
 K^\pm: 
\quad \quad E_K \ge 1.048 \cdot E_\nu \,. 
\label{eq:mu_nu_lab_num} 
\end{eqnarray} 

The corresponding numeric values for the muon--neutrino relations are
$E_\mu \ge 1.342 \cdot E_\nu$ for $\pi^\pm$ decays, and 
$E_\mu \ge 0.048 \cdot E_\nu$ for $K^\pm$ .
It should be noted that - unlike the situation in the cm-frame - the companion  
muon energy can be as low as 5\% of the neutrino energy, 
if the parent  meson is a kaon. 
The difference between pion and kaon decays which  
is related to the quadratic mass ratios $r_i$, 
has an important impact on the energy dependence of
the veto efficiency as discussed below. 
In particular, in addition to satisfying Eq.~\ref{eq:mu_nu_lab},
the muon must have enough energy to penetrate to the detector.  

To veto \dnu, the \dmu\ track must also be relatively near to the neutrino
trajectory.  From the values of the transformed angles $\Theta_\mu, \, \Theta_\nu$ 
we obtain typical distances between the neutrino and companion muon tracks after 
10~km path length of less than 1~m (0.1~m) for neutrino energies above 
1~TeV (10~TeV) if the parent meson was a pion, and less than 10~m (1~m) 
for kaons. High energy atmospheric neutrinos and their companion muons can 
therefore be treated as quasi-aligned given the typical granularity of 
optical modules in a neutrino telescope. 


The production spectrum of atmospheric neutrinos  
from $\pi^\pm\,(K)\rightarrow\mu^\pm +\nu_\mu$ is
obtained from the convolution of the parent meson decay probability with the spectrum
of mesons and the phase space distribution of the neutrinos:
\begin{eqnarray}
\label{M2nu}
{\cal P}_\nu(E_\nu,X)= \hspace{6cm} \\ \nonumber
\int_{E_{\pi,{\rm min}}}^\infty
\left[{B_{\pi\rightarrow\mu\nu}\over E\,(1-r_\pi)}\right]
\left\{{\epsilon_\pi\over E\,X\,\cos(\theta)}\right\}\,\Pi(E,X)\,{\rm d}E\\ \nonumber
 + 
\int_{E_{K,{\rm min}}}^\infty
\left[{B_{K\rightarrow\mu\nu}\over E\,(1-r_K)}\right]
\left\{{\epsilon_K\over E\,X\,\cos(\theta)}\right\}\,
K(E,X)\,{\rm d}E.
\end{eqnarray}
Here ${\cal P}_\nu{\rm d}E_\nu$ is the number of neutrinos ($\nu_\mu+\bar{\nu}_\mu$)
with energy between $E_\nu$ and $E_\nu + {\rm d}E_\nu$
produced per g/cm$^2$ along the direction defined by zenith
angle $\theta$.  $\Pi(E,X)$ and $K(E,X)$ are the
differential energy spectrum of charged pions and kaons at slant depth 
 $X$ (in g/cm$^2$), and $E$ is the energy of the parent pion or kaon.
The factors in curly brackets are the decay probabilities
of pions (kaons)
at vertical depth $X\cos(\theta)$. The factors in square brackets give the 
branching ratios times the normalized distributions
of neutrino energies.
The decay distributions are isotropic in the parent rest frame
(uniform in $\cos\theta_\nu$),
so from Eq.~\ref{eq:Eapprox}, the distributions of neutrino
energy are flat over the kinematically allowed
regions of phase space:
$0\le E_\nu \le E_\pi\,(1\,-\,r_\pi)$ and
$0\le E_\nu \le E_K\,(1\,-\,r_K)$.
The total differential intensity of neutrinos
is obtained by integrating Eq.~\ref{M2nu} over the whole atmosphere
with $E_{\pi,min}$ and $E_{K,min}$ given by Eqs.~\ref{pions} and~\ref{eq:mu_nu_lab_num}.

\begin{figure} 
\vspace{-.5cm}
\begin{center}
\noindent
\includegraphics[height=0.4\textheight]{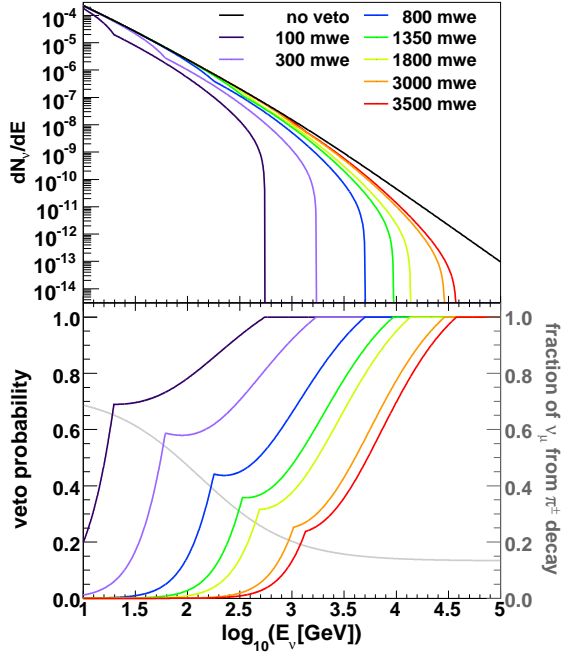}
\end{center}
\vspace{-.5cm}
\caption{Unaccompanied vertical atmospheric \dnu\ flux (upper); probability
of accompaniment (lower) for seven depths.  The grey curve is the fraction of \dnu\
from $\pi$ decay. 
}
\label{fig2}
\end{figure}

The contributions from pions and kaons in Eq.~\ref{M2nu} are identical in form, but the kinematics
are significantly different because of the difference in mass ratios.
The critical energies below which decay is favored over hadronic
re-interactions are also different, $\epsilon_\pi\approx 115$~GeV
and $\epsilon_K\approx 850$~GeV.  As a consequence of these differences the
contribution of pions decreases and kaon decay becomes
the main source of atmospheric
neutrinos at high energy.  (See $\pi$ fraction in Fig.~\ref{fig2}.)

If we require a minimum muon energy at production
so that the muon can penetrate to the detector at slant-depth $X$, then,
in addition to Eq.~\ref{eq:mu_nu_lab}
\begin{equation}
E\;\ge\;E_\nu\,+\,E_{\mu,{\rm min}}(X)
\label{general}
\end{equation}
must also be satisfied.  The latter condition governs until
$E_\nu > E_{\mu,{\rm min}}(X)\times(1-r)/r$, which is 
$E_\nu > 0.75\times E_{\mu,{\rm min}}(X)$ for pion decay but
$E_\nu > 20\times E_{\mu,{\rm min}}(X)$ for kaon decay.
For sufficiently high neutrino energy, the accompanying
muon is guaranteed at depth, but the asymptotic energy
occurs later for the dominant kaon component.

\begin{figure} 
\vspace{-.5cm}
\begin{center}
\noindent
\includegraphics[height=0.4\textheight]{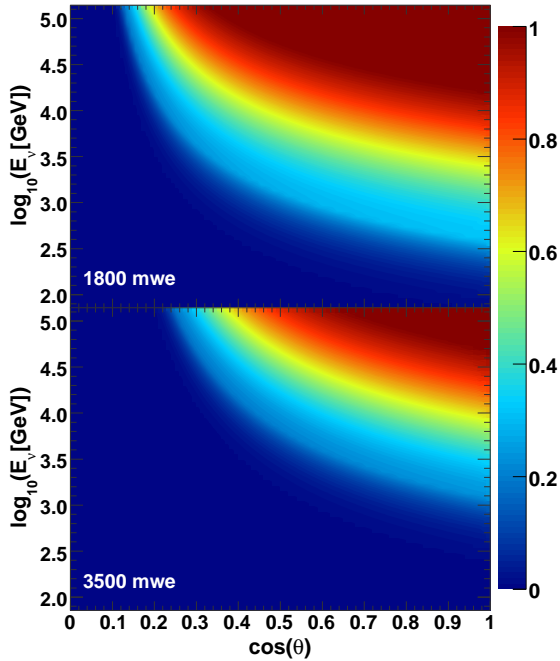}
\end{center}
\vspace{-.5cm}
\caption{Veto probability as a function of zenith angle for 1800 
and 3500 m.w.e. 
}
\label{fig3}
\end{figure}

Fig.~\ref{fig2} illustrates
the behavior of the proposed atmospheric \dnu\ veto as a function
of depth at vertical incidence.  The lower plot shows the probability that
the partner muon reaches various depths as a function of $E_\nu$.
In each case the early onset of veto for the $\pi\rightarrow\mu\nu$ component is
indicated by the shoulder of the curve.
The upper panel shows the remaining flux of atmospheric \dnu\ 
after the \dmu\ veto is applied.  
Fig.~\ref{fig3}
shows how the veto probability depends on zenith angle at two different
depths, 1.8 and 3.5 km.w.e.  The first approximates the center of IceCube at 
2 km in ice, while the second represents the center of a detector
at the NEMO site, which is the deepest candidate location for Km3NeT~\cite{km3net}.

For this illustration we use a simple energy-independent, average relation 
from MMC~\cite{RangeRef} for the 
muon energy at production needed to reach depth $X$:
$$
E_{\mu,{\rm min}}(X)=0.73\,{\rm TeV}\times \left\{
\exp\left[{X/ 2.8\,{\rm km.w.e.}}\right]\,-1\,\right\}.
$$
The \dnu\  veto probability decreases with depth for a given neutrino direction and
with increasing zenith angle at a fixed vertical detector depth.  These effects
are both the result of the increased muon energy needed to penetrate to
the deep detector.  
The dominant kaon
contribution reaches its asymptotic value about a factor 30 higher in
energy than the pion component.

The values in Figs.~\ref{fig2} and~\ref{fig3} 
are obtained from an analytical approximation
applicable for a power law primary cosmic-ray spectrum~\cite{Gaisser}.
For pions
\begin{eqnarray}
\label{Pi-approx}
\Pi(E,X)&=&e^{-(X/\Lambda_\pi)}{Z_{N\pi}\over \lambda_N}N_0(E) \\ \nonumber
  &\times &         \int_0^X\exp {\left[{ X'\over \Lambda_\pi} -{X'\over
  \Lambda_N}\right] }\left ({X'\over X}\right )^{\epsilon_\pi /E\cos\theta}\,
{\rm d}X',
\end{eqnarray}
with a corresponding expression for kaons.  $N_0(E)$ is the power-law
primary spectrum of nucleons evaluated at the pion energy.  Cross coupling
between kaon and pion channels has been neglected as well as 
production of anti-nucleons.  $Z_{N\pi}$ is the spectrum-weighted moment
for pion production, $\lambda_N$ is the nucleon interaction length and
$\Lambda_N,\;\Lambda_\pi,\;\Lambda_K$ are attenuation lengths for nucleons, 
charged pions and charged kaons, respectively.

Eq.~\ref{M2nu} gives the production spectrum of neutrinos
as a function of slant depth $X$ in the atmosphere.  The total 
differential neutrino spectrum requires evaluating
$\int_0^{\rm ground}{\cal P}_\nu(E,X){\rm d}X$.
High-energy muons originate high in the atmosphere, 
so it is a good approximation to extend the upper limit
of this integral to infinity.  The result for the contribution from 
charged pions is
\begin{eqnarray}
\label{nuflux}
&\phi_{\nu,\pi}(E_\nu)=N_0(E_\nu)Z_{N\pi}{\Lambda_\pi\over\lambda_N}
{\xi(E_\nu)\over 1-r_\pi}
\times \int_{z{\rm min}}^\infty{{\rm d}z\over z^{\gamma+2}} \hspace{1cm} \\ \nonumber
&\left[{1\over z+\xi(E_\nu)}-{\Lambda_\pi/\Lambda_N-1\over 2z+\xi(E_\nu)}
+{(\Lambda_\pi/\Lambda_N-1)^2\over 3z+\xi(E_\nu)} + \ldots\right],
\end{eqnarray}
with a similar expression for kaons.  Here $z = E/E_\nu$ and
$z_{\rm min}$ is the greater of $1+\,E_{\mu,{\rm min}}\,/\,E_\nu$ from Eq.~\ref{general}
or $1/(1-r_i)$ (Eq.~\ref{eq:mu_nu_lab}).
Also, $\xi_i(E)=\epsilon_i/E\cos(\theta)$.

The integral in Eq.~\ref{nuflux} can be evaluated analytically in the
limits of low ($E_\nu\cos(\theta)\ll \epsilon_\pi$) 
and high ($E_\nu\cos(\theta)\gg\epsilon_\pi$) energy.
One can then combine the low and high energy limits into
a single approximation,
\begin{equation}
\label{nuflux-approx}
\phi(E_\nu)\approx {N_0(E_\nu)\over 1-Z_{NN}}\Sigma_{i=\pi,K}
\left[{A_i\over 1+B_i/\xi_i(E)}\right],
\end{equation} 
\vspace{-.5cm}
$$
{\rm where}\,\,\,\,\,A_i={Z_{N,i}\over 1-r_i}\,{1\over\gamma+1}\,{1\over(z_{i,min})^{\gamma+1}}
$$
\vspace{-.5cm}
$$
{\rm and}\,\,\,\,\,\,B_i=z_{i,min}\,{\gamma+2\over\gamma+1}\,{\Lambda_i - \Lambda_N\over\Lambda_i\ln(\Lambda_i/\Lambda_N)}.
$$
The expression \ref{nuflux-approx} with numerical values of the parameters
from Ref.~\cite{Gaisser} is used to make Figs.~\ref{fig2}
and~\ref{fig3}.

%

The new opportunity to veto high energy atmospheric muons 
using part of a neutrino telescope \cite{deepcore},
opens the possibility to suppress downward going atmospheric neutrinos.
Extra-terrestrial neutrinos, which are in general not accompanied by 
a muon, will not be discarded by the veto system.
If a downward event starts inside a fiducial volume of the 
detector and has no activity in the outer veto region, and has
sufficiently high energy 
that an atmospheric \dnu\ would be accompanied by its partner muon, then
the neutrino can be classed (with a certain probability) as being of extra-terrestrial
origin. 

So far we have assumed that all neutrinos come from decay of charged kaons and
pions.  We now estimate the contribution of atmospheric \dnu\ from
minor channels and assess their effects on the veto probability.
The $K_{\mu 3}$ semi-leptonic decay has a branching ratio of $0.27\,(K_L^0)$
and $0.033\,(K^\pm)$.  The final state is $\pi,\mu,\nu_\mu$, so the 
minimal kinematic configuration occurs as before when the muon is
backward in the cm system of the decaying kaon, balanced by the
forward moving neutrino and pion.  In this case, however, the
forward momentum is shared by two particles so a lower $E_\nu$ guarantees
the presence of the partner muon for a \dnu .  This channel therefore
does not weaken the veto probability.  At the other extreme are the
\dnu\ from muon decay, in which case there is no muon.
The contribution is small at high energy and can be estimated
as in Ref.~\cite{Lipari}.  For $E_\nu\sim 10$~TeV and zenith
angles $<60^\circ$, this contribution is less than one per mil.

Contributions of prompt neutrinos from charm are more complex
and more difficult to assess.  A model for charm production
that nearly saturates existing limits~\cite{Gelmini} is the
Recombination Quark Parton Model (RQPM)~\cite{Bugaev}.  
The charm contribution to atmospheric \dnu\ is less than 10\% 
for $E_\nu=10$~TeV in this model and crosses over the conventional contribution 
just above $100$~TeV.  Relevant decay channels include
$D^+\rightarrow \bar{K}_0\,\mu^+\,\nu_\mu$~(7\%),
$D^+\rightarrow K^-\,\pi^+\,\mu^+\,\nu_\mu$~(4\%),
and $\Lambda_c^+\rightarrow\Lambda\,\mu^+\,\nu_\mu$~(2\%).
Because the charmed mass is higher, the energy fraction
carried by the muon can be even lower than in the case of the kaon.
On the other hand, the final states involve several particles so the
neutrino energy required to see the partner muon is relatively low
as well.  Given the branching ratios and the upper limits
on charm contribution, unaccompanied prompt \dnu\ should weaken the
veto probability by less than one percent.

We conclude therefore that it should be possible in principle to
veto atmospheric neutrinos with energies in the multi-TeV range and
zenith angles less than $60^\circ$ with an efficiency of
99\% or somewhat better. 
This paves the way for a sensitive and new type of search of neutrinos
from astronomical sources.
The main limitation will  be the
extent to which real detectors can put a lower limit on the energy
of a neutrino interaction that starts in the detector.  
Because of the steep spectrum of atmospheric neutrinos, most
of the interactions will be near the nominal threshold,
thus enhancing the smearing effect of fluctuations.
Detailed simulations of individual detectors, 
including surface air shower arrays, or alternative veto systems at shallow
depths,
are needed
to assess the veto capability in practice.  Such simulations will also
be able to include the case where the \dnu\ is vetoed by a
muon on a different branch
in the accompanying air shower, which enhances the veto probability to some 
extent at high energy.

We thank P.O. Hulth, T. DeYoung, C. Wiebusch and A. Karle for useful
discussions.
T.G. is supported in part by a grant from the
National Science Foundation (ANT-0602679).
E.R. and O.S. are funded by the Deutsche Forschungsgemeinschaft (DFG)
through an Emmy Noether-grant (RE 2262/2-1).
The research of St.S. is supported in part by 
SFB/TR27 `Neutrinos and Beyond'
of the DFG.

\end{document}